# Comprehensive numerical analysis of doping-controlled efficiency in lead-free CsSn$_{1-x}$Ge$_x$I$_3$ perovskite solar cell

Nazmul Hasan[1‡], M. Hussayeen Khan Anik[2‡], Mohammed Mehedi Hasan[3‡], Sharnali Islam[4,5 Ψ], Alamgir Kabir[6] Ψ

[1]Department of Electrical and Computer Engineering, University of Rochester, Rochester- 14627, USA, nzharis97@gmail.com,

[2] Department of Electrical and Computer Engineering, University of Delaware, Newark, DE 19716, USA. mhkanik@udel.edu,

[3]Department of Theoretical Physics, University of Dhaka, Dhaka 1000, Bangladesh, mehedi.tpdu@gmail.com

[4]Department of Electrical and Electronic Engineering, University of Dhaka, Dhaka 1000, Bangladesh,

[5]Semiconductor Technology Research Centre, University of Dhaka, Dhaka 1000, Bangladesh
sharnali.eee@du.ac.bd

[6]Department of Physics, University of Dhaka, Dhaka 1000, Bangladesh, alamgir.kabir@du.ac.bd

[‡]Authors contributed equally.

[Ψ]To whom it corresponds: sharnali.eee@du.ac.bd, alamgir.kabir@du.ac.bd

*Abstract:* One effective way to prevent toxicity and improve the stability of materials for photovoltaic applications is to exclude lead and organic molecules from perovskite materials. Specifically, the CsSn$_{1-x}$Ge$_x$I$_3$ appears to be a promising contender; nonetheless, it requires optimization, particularly bandgap tuning by doping concentration modifications. In this study, density functional theory (DFT) was employed to comprehensively analyze the electronic properties of CsSn$_{1-x}$Ge$_x$I$_3$ that influenced light-matter interactions tuning of the perovskite materials by varying composition in B site atoms. We use the solar cell capacitance (SCAPS-1D) simulator to compute device performance; however, it computes the absorption spectrum using a simplified mathematical function that approximates the actual spectrum. To achieve a quantum-mechanical level of accuracy DFT extracted parameters like absorption spectra and bandgap were fed into SCAPS-1D. We find that increasing the Ge concentration leads to a higher bandgap and improved absorption profile, thereby enhancing solar energy conversion efficiency. Thermal and field distribution analyses were also done for the optimized device through a finite-difference time-domain (FDTD) framework. By optimizing the absorber layer with a 75% Ge concentration, we achieve a remarkable PCE of 23.80%. Our findings guide future research in designing high-performance non-leaded halide PSCs, paving the way for low-cost, stable, and highly efficient solar cells through atomic doping-tuned perovskite absorber layers.

*Index terms*: Inorganic Perovskites, Solar Cell Efficiency, Density Functional Theory, *SCAPS-1D, FDTD*, Doping, Optical Properties, Thermal Stability, Temperature Distribution.

# INTRODUCTION

The development of highly efficient and stable solar cells is critical for achieving a sustainable and low-carbon energy future. It is evident that solar energy being copious, low-cost, and eco-friendly requires rapid development of efficient solar energy harvesting devices i.e., solar cells. Furthermore, technological maturity nowadays can utilize thermoelectric properties of materials to generate electrical power. The rapid boost in power conversion efficiencies of perovskites-based solar cells has triggered enormous investigations toward understanding the fundamental properties of these materials. Halide perovskites have recently emerged as a promising class of materials for next-generation optoelectronics due to their excellent photophysical properties, including high light absorption coefficient, long carrier diffusion length, and tunable bandgap. Among the different types of halide perovskites, non-leaded ones have attracted particular attention due to their lower toxicity and environmental impact compared to lead-containing counterparts. Despite significant progress in improving the efficiency of non-leaded halide PSCs, achieving high power conversion efficiency (PCE) with long-term stability remains a major challenge.

Generally, perovskite materials exhibit fascinating optoelectronic properties that stem from their unique crystal structure [$ABX_3$] which consists of an in/organic cation [A], and an inorganic framework composed of metal-halide [$BX_6$] octahedra [1]. The crystal structure arrangement of perovskites creates a favorable electronic band configuration for unique light-matter interaction properties resulting in efficient, wide-range light-absorption and excellent charge carrier transport properties including efficient exciton generation, and effective charge separation which in turn contribute to the high efficiency of PSCs [2]. However, the efficiency of solar cells depends on the capability of perovskite to effectively convert incident light into usable electrical energy. The light-matter interaction in PSCs begins with the absorption of photons (incident light) by the perovskite absorber layer and generates an exciton—a bound electron-hole pair [3]. Excitons in perovskite materials are characterized by strong binding energies, resulting in relatively long exciton diffusion lengths that can travel relatively large distances before recombining, which is crucial for efficient charge separation [4]. The energy level alignment at the interfaces between the perovskite layer with electron-transporting material (ETL), and hole-transporting material (HTL) materials in solar cell heterojunction facilitates the separation of the charges, allowing electrons – holes conduction [5]. To enhance exciton dynamics and overall efficiency in solar cell devices, researchers have investigated the incorporation of additional materials in interface transport layers, such as electron or hole selective contacts, to optimize charge extraction, minimize recombination losses, and improve charge transport pathways [6].

By harnessing the unique and tunable structural physics-based light-matter interaction properties of perovskite materials, researchers strive to enhance solar cell efficiency, stability, and scalability. From that point forward, optimizing the interface between the electron-transporting layer (ETL) and hole-transporting layer (HTL) in a proposed inorganic light-harvesting Bi-iodide PSCs has demonstrated impressive power conversion efficiency (PCE) of approximately 14% [7]. Interestingly, it has been observed that defects within the perovskite layer have a more pronounced impact on the physical properties and device performance compared to defects in the ETL and HTL materials. A comprehensive density functional theory and experimental study have reported more than 17% PCE upon tuning the optoelectronic properties of ETL by varying thickness and atomic doping with 20 nm scaled thinner and 4% Sn-doped $TiO_2$ in the ETL layer [8]. Also, engineering with HTL along with varying absorber layer thickness demonstrates to enhance the device efficiency from 15% to 20.49% by tuning radiative recombination rates [9]. Following these findings, solar cell researchers are investigating more to gain better device efficiency by engineering ETL/HTL-Perovskite materials and some recent studies found a maximum of 17%-22% PCE with varying materials compositions [10], [11], [12], [13], [14], [15]. However, doping concentration in a halogen site composed of Br-I within a Sn-based organic cubic perovskite shows a significant PCE increment from 6.71% (pure) to 20.72% (mixed halide) which indicates that optimal halogen mixing also can be considered a way to tune the device efficiency [16]. Another novel approach utilizing the traditional $ABX_3$ perovskites-on-Si tandem solar cell heterojunction has been implemented, resulting in significant enhancements in practical device efficiency [17]. Remarkably, the organic $CH_3NH_3SnI_3$ exhibited an excellent PCE of 33.41%, while the inorganic lead-free $CsSnI_3$ achieved a noteworthy PCE of 31.09% through different solar cell device optimization [18], [19]. Both organic and inorganic PSCs demonstrate that optimizing contact materials, along with engineering the thickness of ETL-HTL variations, can significantly enhance the light-matter interaction-driven electrical efficiency in solar cell devices. In comparison, their standalone tandem devices yielded PCE values of 20.01% and 7.45% respectively. Solar energy-based efficient light energy harvesting materials and device researchers are still exploring to find noble materials for having stable and highly yielded PCE solar cell devices and appeared with numerous compositions to get desired device efficiency, i.e., transition metal composed PSC [20], stacked 2D materials-PSC [21], triple cation mixed halide perovskites [22], Moiré perovskites [23], perovskite-metal-dielectric solar cell [24], and so on.

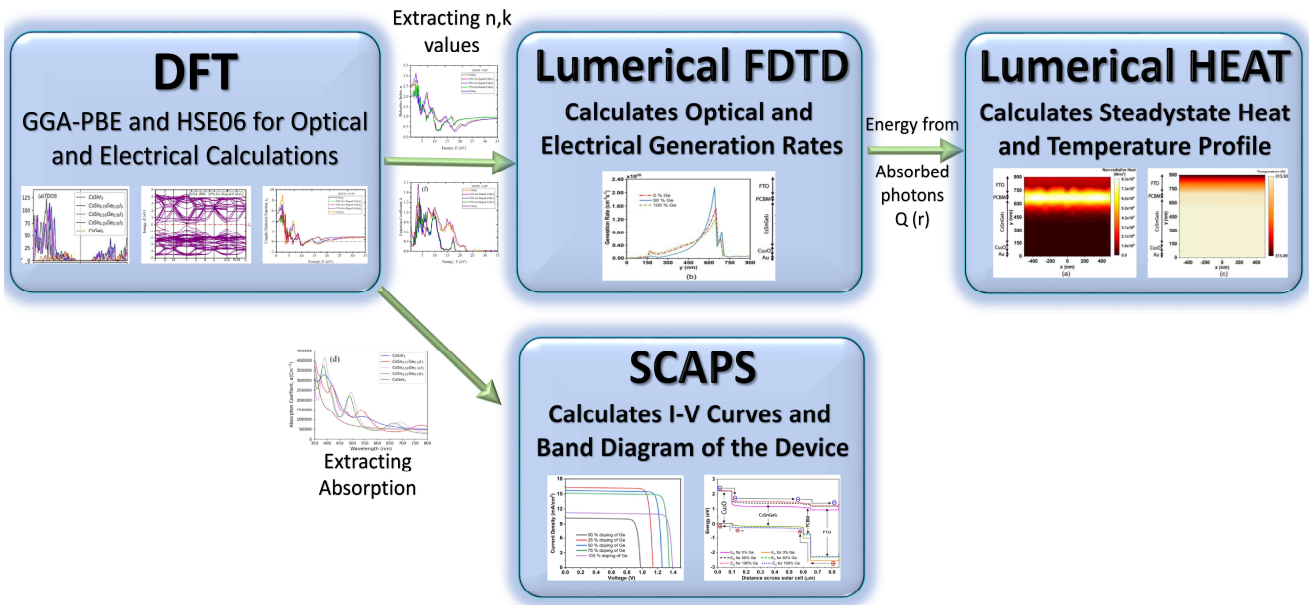

**Fig. 1.** Workflow of the current study combining the multimodal frameworks.

Though numerous studies on materials-specific device performance are now available in the community, a comprehensive understanding of the interplay between the structural, electronic, and optical properties of the absorber layer and their impact on the device performance is still lacking. In this work, we use first-principles DFT calculations to extract the electronic and optical properties of non-leaded halide $CsSnI_3$ PSCs through absorber layer variation by atomic doping of Ge in the inorganic framework's B-site. Next, the PSC's performance metrics were derived by widely used SCAPS-1D software [25], [26], [27]. Even though SCAPS-1D has been quite successful in predicting PSC performance, the built-in calculation of the absorption spectrum is based on the combination of power laws, which oversimplifies the result [8], [28]. This can compromise the precision of SCAPS-1D's prediction and lead to inaccurate PCEs. To overcome this, instead of the in-built absorption spectra, the DFT-extracted spectra and the bandgap were used in the SCAPS-1D simulator in this work. This investigation encompassed carrier dynamics within the solar cell, considering factors such as trap-assisted recombination, the space-charge region, and interface recombination. Finally, we employ the FDTD method to model light absorption and scattering properties of the absorber layer. Then we couple the optic module to the thermal module to calculate the distribution of non-radiative heat and the temperature gradient in each layer of the solar cell. Hence, merging the DFT calculated parameters with the SCAPS-1D simulator ensures a highly accurate method to calculate PCE, and the overall flow of this work is depicted in **Fig. 1**. Through the systematic optimization of the absorber layer, we achieve a PCE over 23%, which is among one of the highest reported for non-leaded standalone Sn-halide PSCs with typical ABX$_3$-solar cell heterojunctions so far. Our results provide a detailed understanding of the interplay between the structural, electronic, and optical properties of the absorber layer and their impact on the efficiency and stability of non-leaded halide PSCs.

## NUMERICAL MODELLING AND METHODS

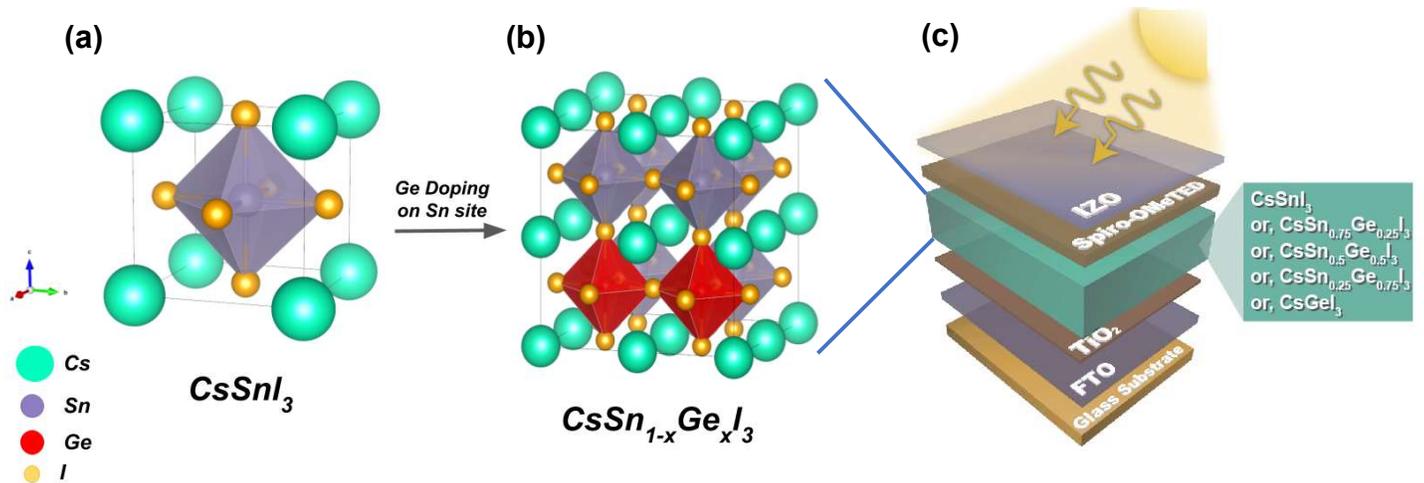

**Fig. 2.** DFT optimized crystal structure for (a) CsSnI$_3$ cubic Pm$\bar{3}$m intrinsic unit cell, (b) 2×2×2 supercell for (x = 0.25, 0.50, 0.75) Ge doping in CsSnI$_3$ perovskite, and (c) Modelled device structure.

*Density Functional Theory (DFT) simulations:*

A standard CsSnI$_3$ perovskite crystal unit cell was initially chosen for structural modeling, followed by its transformation into a 2×2×2 supercell of CsSnI$_3$ perovskite crystal as in **Fig. 2(a) and (b)**. Substitution of Sn with Ge atoms at 25%, 50%, 75%, and 100% concentrations was performed for first-principle DFT calculations. DFT [29], [30] calculations were carried out using the generalized gradient approximation (GGA) [31] and considering the projector-augmented wave (PAW) [32] method, as conducted in the *VASP* code [33]. The plane wave cutoff energy was set to 500 eV. For the electronic structure, both the PBE [31] functional and the HSE06 [34], [35] hybrid functional (75% Hatree-Fock exchange-correlation) are employed, as the PBE functional slightly underestimates the bandgap. A k-mesh of 6×6×6 grid Monkhorst pack [36] is utilized for pure structures, while a 3×3×3 mesh is used for doping the Ge atom in the Sn site of the $CsSnI_3$ supercell. Complete electronic relaxation of the ionic coordinates, shape, and volume of the unit cell was continued until the self-consistent total energy was adjusted to $10^{-8}$ eV per atom. The atomic coordinates of the structures were relaxed until the total force on each atom was less than 0.001 eV. Since electronic structures are used to analyze optical properties, we implemented HSE06 for optical calculations.

*Device Structure:*

In this work, lead-free, environmentally friendly $CsSn_{1-x}Ge_xI_3$ (x = 0.00, 0.25, 0.50, 0.75, and 1.0) is employed with different doping levels of Ge as the absorber material in the PSC with an n-i-p configuration where PCBM is chosen as ETL, and $Cu_2O$ is selected as HTL. The structure is accompanied by Fluorine-doped Tin Oxide (FTO) as a transparent glass oxide placed at the top layer in the PSC which acts as a cathode whereas Au is considered as back metal anode contact. FTO is used as the cathode layer because of its strong optical transparency, high substrate adhesion, and chemical inertness, whereas Au is employed as the back metal contact because of its improved electrical conductivity, stability, and greater compatibility with perovskite materials [37], [38]. FTO and Au are assumed to have work functions of 4.4 eV and 5.45 eV, respectively [39]. **Fig. 2(c)** displays the schematic representation of the multilayer PSC structure utilized for device modeling in this work. **Table S1** in the supplement provides a summary of the key electrical input parameters that were employed in the simulation. The layer thicknesses for FTO, PCBM, $CsSn_{1-x}Ge_xI_3$, $Cu_2O$, and Au contact are 200 nm, 50 nm, 500 nm, 100 nm, and 60 nm respectively. The design provides contact defects between both HTL/perovskite and ETL/perovskite interfaces. An unbiased Gaussian pattern of defects with a characteristic energy of 0.1 eV and 0.6 eV above the VB, respectively, has been produced within the perovskite layer. The ETL/perovskite and perovskite/HTL interface defects are described as neutral single defects with an energy of 0.6 eV above the VB.

*Electrical and Thermal Modelling:*

We investigated the effect of different Ge doping levels (x = 0.00, 0.25, 0.50, 0.75, and 1.00) in $CsSn_{1-x}Ge_xI_3$ on the performance of the PSC device using two optoelectronic simulators, SCAPS-1D, and FDTD-based Ansys Lumerical 2022 software. To study device performance varying with Ge concentration in the absorption layer, the absorption spectra and band gap of the perovskite absorber layer $Cs(SnGe)I_3$ are acquired via DFT calculations. These parameters are then fed into SCAPS-1D to determine the current voltage characteristics of the physical solar cell. By linking the basic photovoltaic equations for each layer, which are controlled by user-defined design and material factors, SCAPS uses photovoltaic models to emulate any form of solar cell response. Due to the high dielectric constant of the perovskite framework, incoming photons can be quickly separated into free charge carriers after being absorbed by them [40]. The Poisson equation, carrier continuity equations, and drift-diffusion charge transport model equations are the three guiding principles that determine the properties of carriers in photovoltaic devices. SCAPS coordinates the device simulations using these core models in addition to other photovoltaic equations including recombination losses and optical absorption (details can be found in supplementary Section 2).

In the final stage of our device modeling, we looked at the electric field distribution, generation rate, non-radiative heat distribution, and temperature profile of our PSC using the FDTD approach in Ansys Lumerical. In our FDTD simulation, we utilized absorber refractive index (n) and extinction coefficient (k) data derived from hybrid HSE06 functional-based DFT optical calculations. We employed Palik data for Au metal and drew upon the n and k values obtained from prior research studies for materials such as $Cu_2O$, FTO, and PCBM [41], [42], [43], [44]. Following these parameters, a 3D model of our proposed PSC is built for the simulation process. For the Y and Z directions of the solar cell, perfect-matched layer (PML) boundary conditions are employed, whereas periodic boundary conditions are used for the X direction. The rate of electron-hole pair generation can be obtained from electric field orientation ($\vec{E}_{op}$) and imaginary part of complex permittivity ($I(\vec{r}, \omega)$) [45].

$$G(\vec{r}) = \int -\frac{\pi}{h} \left| \vec{E}_{op}(\vec{r}, \omega) \right|^2 * I\{\epsilon(\vec{r}, \omega)\} \qquad (1)$$

Here, $G(\vec{r})$ represents the electron-hole pair generation across the solar cell and $h$ denotes the Planck's constant.

We have looked at the nonradiative heat distribution and temperature profile in Lumerical HEAT, which are significantly connected with the optical and electrical systems of solar cells, in addition to generation rate and electric field calculation in Lumerical FDTD. The solution to the differential equation of steady-state conduction with internal heat production determines the temperature distribution [46].

$$-k\nabla^2 T + Q = \rho_p C_P \frac{dT}{dt} \qquad (2)$$

where k represents the temperature-dependent thermal conductivity of a material, $C_P$ represents its specific heat, and $\rho_p$ represents its density. Q defines the amount of net energy absorbed by the surface of a sample exposed to solar radiation. The conventional AM 1.5 G spectrum of solar radiation at 300 K temperature is used to enlighten the structure from the top at normal incidence. Simulations were performed in the wavelength range of 300 nm to 2000 nm. The incident monochromatic photon's frequency is set at $10^{18}$ s$^{-1}$. For different absorber materials, the solar cells' series and shunt resistances are tuned at 4200 Ohms/cm² and 1 Ohm/cm², respectively. (See section **S2**, in the supplementary information for Device Modelling and Methods.)

## Physical Chemistry of the CsSnI₃ Perovskites:

### Structural Properties

Structural relaxation of the energetically stable phases for the considered Ge-doped and undoped CsSnI₃ are obtained through first-principle calculations leveraging density functional theory. Geometrically optimized systems are crystallized in the cubic lattice units with Pm$\bar{3}$m (221) space group for undoped crystals whereas both PBE and hybrid HSE06 functional relaxation results in Fm$\bar{3}$m (225) symmetrical cubic phases for doped structures. Essential parameters for studying the structural dynamics and stability of the geometrically optimized perovskite phases are listed in **Table 1**. PBE and HSE06 functionals show distinct behavior in the change of the structural behaviors for the cubic halide perovskites. Considering lattice constants and unit cell volume, HSE06 functionals lead to the PBE functional, where an increasing trend is observed for both parameters with changing the Ge doping concentration ranging from 25% to 100%. These variations lead to quantum-confined electronic properties variation in the perovskites which is analyzed in the later section. In the perovskite materials, octahedral framework [BX$_6$] significantly influences to manipulation of the optoelectronic behavior of the materials leading to device efficiency change [47]. Increasing the concentration of Ge in CsSnI$_3$ perovskites results in increasing the octahedral bonding behaviors by increasing the interatomic bond lengths as observed in **Table 1**. Octahedral distortion plays a strong role in perovskites' structural stability as well and hence the tolerance factor (τ) is accordingly determined using the following relations [48], [49]:

$$\tau = 0.707 \times \frac{R_A+R_X}{R_B+R_X} \text{ [for pure stuctures]} = 0.707 \times \frac{R_A+R_X}{\left(\frac{R_{Sn}+R_{Ge}}{2}\right)+R_X} \text{ [for Ge doped structures]} = 0.707 \times \frac{R_A+R_X}{[(1-x)R_{Sn}+xR_{Ge}]+R_X} \text{ [for varying Ge concentration)  (for cubic perovskites' stability } 0.8<\tau<1) \quad (3)$$

Following the criterion as determined by the relations (3), all the studied perovskites demonstrate their stable nature in ambient conditions indicating suitability for device fabrication processes as can be observed from **Table 1**.

Table 1: Geometrically relaxed structural parameters, electronic band gaps, and static optical indexes for the pristine and Ge doped CsSnI$_3$ cubic perovskites.

| Perovskite | CsSnI$_3$ | CsSn$_{0.75}$Ge$_{0.25}$I$_3$ | CsSn$_{0.50}$Ge$_{0.50}$I$_3$ | CsSn$_{0.25}$Ge$_{0.75}$I$_3$ | CsGeI$_3$ |
|---|---|---|---|---|---|
| Geometry-optimized lattice phase | Pm$\bar{3}$m | Fm$\bar{3}$m | Fm$\bar{3}$m | Fm$\bar{3}$m | Pm$\bar{3}$m |
| Optimized lattice Constants (Å) | 6.17606 | 12.41119 | 12.25843 | 12.13836 | 5.97045 |
| Relaxed Volume (Å$^3$) | 235.58 | 1915.398748 | 1850.027869 Å | 1788.461237 Å | 212.823825 |
| Octahedral Interatomic Bonds | I-Sn: 3.08Å phi(I-Sn-I): 90° | I-Sn: 3.11 Å Ge-I: 3.02Å phi(I-Sn/Ge-I): 90° | Ge-I: 2.99 Å Sn-I: 3.14 Å phi(I-Sn/Ge-I): 90° | Sn-I: 3.11 Å I-Ge: 3.03 Å phi(I-Sn/Ge-I): 90° | Ge-I: 2.98 Å phi(I-Ge-I): 90° |
| Tolerance Factor | 0.96 | 0.98 | 0.98 | 1.00 | 0.99 |
| E$_g$ | 0.22 (PBE) 1.331 (HSE06) [This study] 0.69 (HSE06) [50] 1.3 [51] | 0.25(PBE) 1.483 (HSE06) | 0.41(PBE) 1.695 (HSE06) | 0.52(PBE) 1.840 (HSE06) | 0.53(PBE) 1.927 (HSE06) 0.87 (HSE06) [50] 1.6 [52] |
| ε$_1$(0) | 5.47 | 3.80 | 3.89 | 3.98 | 5.95 |
| n(0) | 2.50 | 1.99 | 2.01 | 2.03 | 2.63 |

*Electronic Properties*

According to quantum many-body physics, confined structural dynamics significantly alter the electronic configuration which results in different light-matter interaction behaviors. Based on the electronic behavior of the materials optical properties differ among insulators, semiconductors, and metals. Hence, electronic band structures and density of states profiles for the considered pristine and Ge doped $CsSnI_3$ are evaluated to deepen the understanding of the constituent elements' behavior. All the electronic properties are calculated along the high symmetry points from the first Brillouin Zone of the considered perovskite phases and represented in **Fig. 3 & 4**.

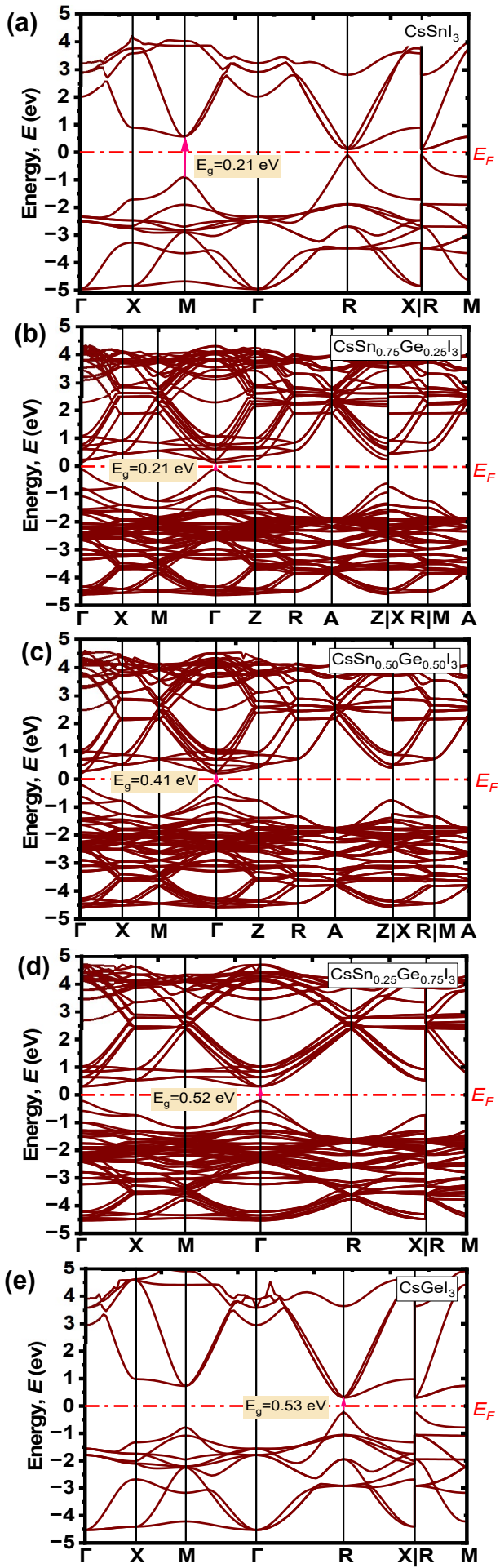
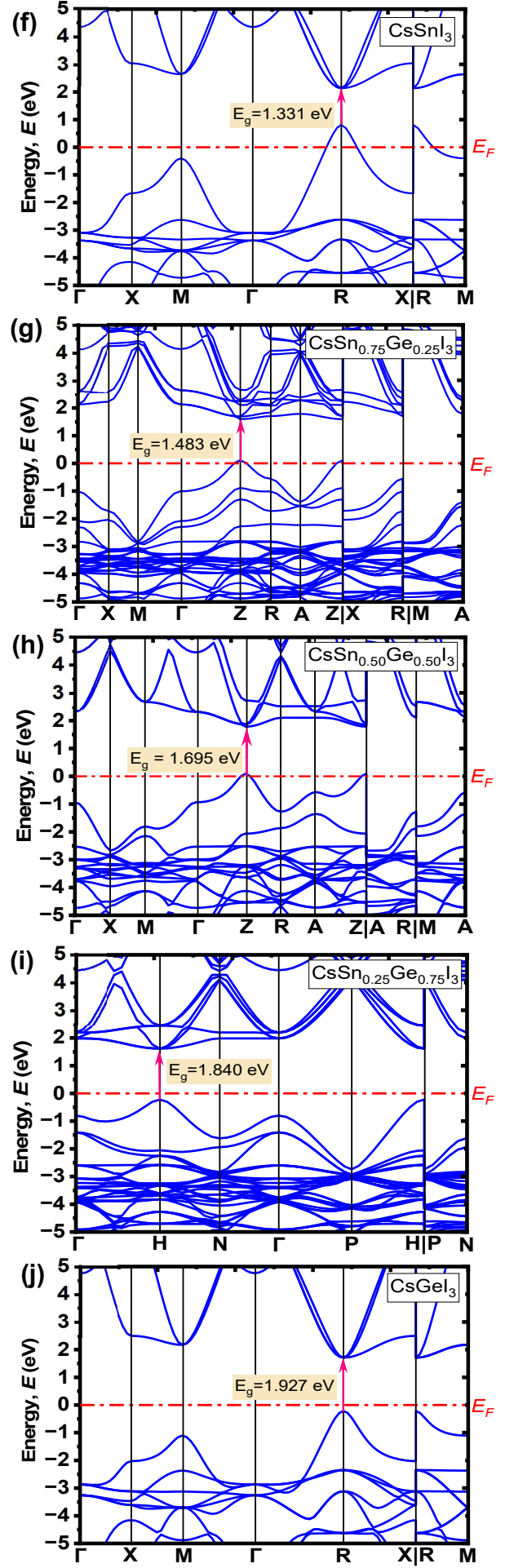

**Fig. 3.** Band structure profiles for: CsSnI$_3$, CsSn$_{0.75}$Ge$_{0.25}$I$_3$, CsSn$_{0.50}$Ge$_{0.50}$I$_3$, CsSn$_{0.75}$Ge$_{0.25}$I$_3$, and CsGeI$_3$. Left panel (a-e) represents GGA-PBE functional-based calculations, Right panel (f-j) represents HSE06 functional-based calculations.

GGA-PBE and HSE06 functional in Density Functional Theory simulations extracted the electronic band structure for the considered CsSnI$_3$ perovskites are portrayed in **Fig. 3**. Comparing the band configurations determined by the two functionals, variation in band gap values with the concentration of Ge in CsSnI$_3$ follows an increasing trend and all compounds are exhibiting direct bandgap semiconducting nature. Perovskites CsSnI$_3$ and CsGeI$_3$ exhibit bandgap values of 0.22 eV (1.33 eV) and 0.53 eV (1.93 eV) respectively in the PBE (HSE06) method at path-R. In **Fig. 3(a-e)**, PBE-based band structure profiles demonstrate a shift in high symmetry K-path from R to Γ for having the direct band gaps for the Ge-doped CsSnI$_3$ perovskites. We have also calculate the band structure of the materials by using CASTEP code (Please see **Fig. S1 in the supplementary information**) and the obtained results are consistent with the one obtained from VASP. Increasing the concentration of Ge in B site by x = 0.25, 0.50, and 0.75 respectively leads to a shift of conduction band minima (CBM) to the higher energy region. GGA-PBE executed band structure profiles with no impurity and defect states indicate the competency of considered CsSn$_{1-x}$Ge$_x$I$_3$ perovskites for potential photovoltaic and optoelectronic device applications. Due to the inadequate consideration of electronic correlation by commonly employed PBE functional, bandgaps are often underestimated. To have a better accuracy in band structures and consequent physical properties for the considered perovskites hybrid functional HSE06 was employed that correspond better with experiments [53] and evaluated band structures are illustrated in **Fig. 3(f-j)**.

HSE06 extracted band gap values also demonstrate the perovskites' semiconducting nature through direct band gap values ranging from 1.3 eV to 1.9 eV intrinsically. The band gap values of CsSnI$_3$ and Ge-doped CSSnI$_3$ materials exhibited a rising trend, in consistent with the trend of PBE executed values as exhibited in **Fig. 3**. The band gap values were 1.3 eV for CsSnI$_3$, 1.48 eV for CsSn$_{0.75}$Ge$_{0.25}$I$_3$, 1.69 eV for CsSn$_{0.50}$Ge$_{0.50}$I$_3$, 1.84 eV for CsSn$_{0.25}$Ge$_{0.75}$I$_3$, and the highest value of 1.93 eV for CsGeI$_3$. Notably, high symmetry points of originating the band gap for the phases are shifted from R point to Z point with the increasing concentration of Ge in the doped perovskites. For the CsSn$_{0.25}$Ge$_{0.75}$I$_3$, high symmetry points of band structure alter with band gap originating k-path shifts from Z to H in the crystal system obtained after geometry optimization. The studied band structure profiles are analyzed within the energy scale of -5eV to 5 eV for the considered perovskites in both PBE and HSE06 approaches. For all the compounds, increasing the concentration of Ge results in an upward shift of conduction band leading to the increment in band gaps.

From **Fig. 3**, it is apparent that substituting Sn with Ge at higher concentrations in the B site results in increased electronic bandgaps compared to the pristine state. The conduction band minima (CBM) shifts to higher energy levels, while the valence band maxima (VBM) remain at the same energy level. The rise in bandgap values is attributed to the decreased atomic radius of the dopant atom Ge ($4s^24p^2$, 0.137 nm) compared to Sn ($5s^25p^2$, 0.162 nm), influencing the interatomic distance and subsequently the binding strength of valence electrons. A reduced interatomic distance enhances the binding forces of valence electrons, requiring more energy to excite them into the conduction band. This phenomenon is reflected in the increasing trend of the energy bandgap, as depicted in **Fig. 3**. Parabolic band approximation at the band edge profiles can. The broadening of the parabolic edge at both the conduction band minimum (CBM) and valence band maximum (VBM) due to Ge doping indicates an increase in effective mass results in a reduction in electron mobility, as supported by the observed increasing trend in the required excitation energy ($E_g$) (**Fig. S2 in the supplementary information**). Moreover, band-edge structure in perovskites is intricately governed by the interaction of atomic orbitals that results in shaping the optoelectronic properties of the materials. Understanding the influence of the band-edge on the optoelectronic properties of CsSn$_{1-x}$Ge$_x$I$_3$ perovskites involves establishing correlations between band-edge electronic structures and states, along with considerations of the Urbach trail and exciton binding energy is useful to project the optical behavior of the material. The partial electron density of states, as projected onto atoms and orbitals in **Fig. 4**, indicates that the origin of valence bands is predominantly from the s-orbitals of Ge/Sn and the p-orbitals of the I atom. Conversely, the conduction band is mainly associated with the p-orbitals of the B site atoms. When observing **Fig. S2** (in supplementary) and **4**, it becomes apparent that as the energy of electronic states approaches the band gap energy, the electron density of states becomes more concentrated in the band edge region that results in a tail of electronic states with energies just below the band edge. The presence of localized states within the band gap of all perovskite materials contributes to the widening of the absorption spectrum near the band edge, attributed to the Urbach tail effect [54]. This effect is crucial in understanding the impact of thermal defects, leading to absorption edge broadening and electronic structure splitting near the band-edges. These observations carry significant implications for various applications, including light-emitting diodes, lasers, and spintronics.

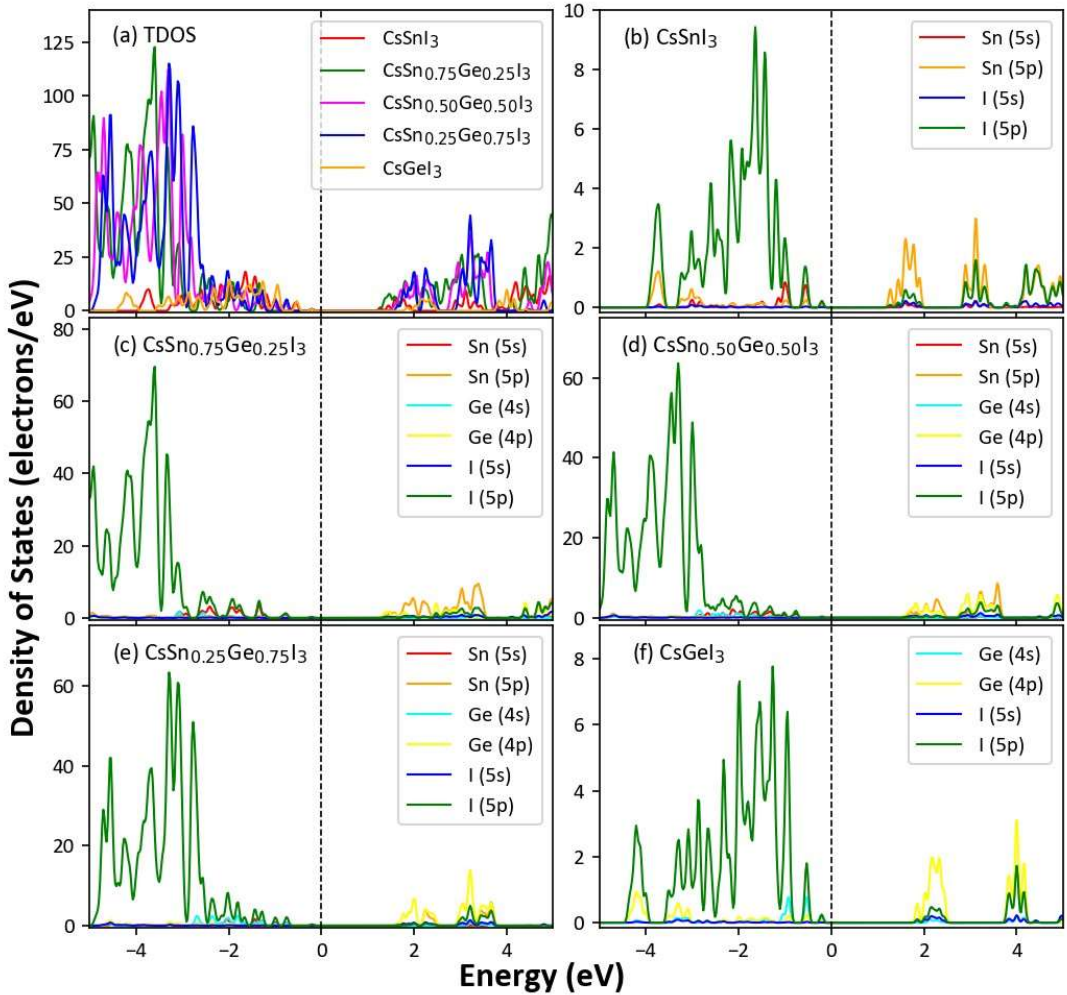

**Fig. 4.** HSE06 employed (a) total density of states (TDOS) for all compounds; partial density of states (PDOS) for- (b) $CsSnI_3$, (c) $CsSn_{0.75}Ge_{0.25}I_3$, (d) $CsSn_{0.50}Ge_{0.50}I_3$, (e) $CsSn_{0.75}Ge_{0.25}I_3$, and (f) $CsGeI_3$.

Describing accurate electron density of states with higher orbitals (d and f) is thoroughly proven by HSE06 hybrid functional within DFT calculations. To examine the atomic and orbital contribution in the perovskites' optoelectronic properties, Fig. 4 represents the DOS profiles from -5eV to 5eV energy range with respect to Fermi Energy ($E_F$) at 0 eV. TDOS of all compounds is demonstrated in **Fig. 4(a)**, where the gap between VBM and CBM is widening with increasing the concentration of Ge doping in $CsSnI_3$. The incorporation of Ge in the B site with Sn in the CsSnI$_3$ perovskites shows that the electron density of states is increased both in the conduction band and valence band energies as observed from the total density of states (TDOS) profiles for the studied perovskites in **Fig. 4(a)**. However, in the deep energy region of valence bands, the CsSn$_{0.75}$Ge$_{0.25}$I$_3$ compound shows a higher magnitude in electron density states whereas in the conduction band, most of the higher energy region is dominated by a 75% Ge doped phase. The electron density of states for the considered perovskites reaches the highest magnitude of 123 states/eV at -3.75 eV in the valence band for x = 0.25 Ge concentration and 40.5 states/eV at 3.12 eV in the conduction band for x = 0.75 Ge doped compound. Compared to the 25% (x = 0.25) Ge-doped compound, 50% (x = 0.50) Ge-doped perovskite reaches to 105.5 states/eV at -3.65 eV (valence band) and 85 states/eV at 3.25 eV (conduction band) while 75% (x = 0.75) doping results to 123 states/eV at -3.75 eV (valence band) and 40.5 states/eV at 3.12 eV in the conduction band.

Near the Fermi region, constructing valence band maxima is mainly contributed by I-5p orbital with a least density state of Sn-5s and Ge-4s orbital electrons as can be observed from the orbital projected partial density of states plots in **Fig. 4(b-f)**. In the deep valence band energies, the appearance of s-orbital contribution by Sn and Ge is visible with the dominance of the I atom's p orbital electrons contribution. Increment in the Ge concentration leads to a rise in the electron states for Ge-p orbital in the lower energy region of the valence band except for x = 0.75 Ge doped phase where Ge's s orbital rose instead of its p orbital contribution. As we go to a higher energy region from the Fermi level, a mixing occurs among dominant Sn-5p/Ge-4p, I-5s5p, and Cs-5p orbitals. As shown in **Fig. 4 (b)**, the top of the CsSnI$_3$ valence band is principally composed of the hybrid states of Sn 5s, Sn 5p, and I 5p, and the bottom of the conduction band contributed by I 5s, I 5p, and Sn 5p states. The top of the CsSn$_{0.75}$Ge$_{0.25}$I$_3$ valence band is mainly contributed by the hybrid states of Sn 5s, Sn 5p, and I 5p, and the bottom of the conduction band by I 5s, I 5p, Ge 4p, and Sn 5p states. Furthermore, the top of the valence band of both CsSn$_{0.50}$Ge$_{0.50}$I$_3$ and CsSn$_{0.25}$Ge$_{0.75}$I$_3$ is dominated by the hybrid states of Sn 5s, Ge 4s, and I 5p, and the bottom of the conduction band by I 5s, I 5p, Ge 4p, and Sn 5p states in **Fig. 4 (d)-(e)**. The only difference is that the CBM shifted to slightly higher values for the CsSn$_{0.25}$Ge$_{0.75}$I$_3$ than for the CsSn$_{0.50}$Ge$_{0.50}$I$_3$ compound. Besides, the top of the CsGeI$_3$ valence band is largely contributed by Ge 5s, Ge 5p, and I 5p orbitals, and the bottom of the conduction band by I 5s, I 5p, and Ge 4p orbitals which are illustrated in **Fig. 4 (f)**. Above all, it is noticeable that the 4s and 4p orbitals of Ge influence the VBM and CBM of $CsSnI_3$.

*Light-matter Interactions*

All the optical properties (absorption coefficient, dielectric constant, conductivity, refractive index, reflectivity, and loss function) for the pristine and Ge-doped $CsSnI_3$ perovskites are studied from 0 - 35eV energy range. The physical properties of materials significantly depend on the energy-dependent dielectric function. The complex dielectric spectra have been described by its two components: real dielectric constant and imaginary dielectric constant following the relation, $\varepsilon(\omega) = \varepsilon_1(\omega) + i\varepsilon_2(\omega)$. The real part of the dielectric function describes the materials' energy storage potentiality with polarization behavior determining the absorption spectra throughout the energy spectrum whereas the imaginary part demonstrates the energy dissipation ability of the material. Since the complex dielectric function lowers the charge-carrier recombination rate and in turn improves the device efficacy, higher dielectric constant values are desired to have higher-quality materials for optoelectronic device applications. Fig. 5 (a, b) illustrates the two components of the complex dielectric function for the pure and Ge-contained CsSnI3 perovskites. The highest magnitudes in $\varepsilon_1(\omega)$ for the considered compounds appeared to be 7.28, 5.35, 6.32, 6.43, 8.98 for $CsSnI_3$, $CsSn_{0.75}Ge_{0.25}I_3$, $CsSn_{0.50}Ge_{0.50}I_3$, $CsSn_{0.25}Ge_{0.75}I_3$ and $CsGeI_3$ perovskites respectively. Besides, static dielectric constants are found in a range of 3.80 to 5.95. After reaching the highest peaks, all real parts of dielectrics fall sharply to negative values and reach almost unity at higher energy regions. However, the imaginary part of dielectric $\varepsilon_2(\omega)$ influenced by the density of states and occupied and unoccupied phases' momentum shows the threshold values found at the closest region of 1.2 eV photon energy. Higher positive dielectric behavior indicates the studied perovskites are good quality materials to have better efficiency in device applications. Peak behaviors in the imaginary dielectric function influence the absorption profile shaping which is identical to **Fig. 5(b)** and **5(c)**.

Photon-energy and wavelength-dependent optical absorption profiles for the considered perovskites in this study are represented in Fig. **5(b)** and **5(c)** respectively. The threshold energy for increment in absorption spectra is found near 1.25 eV which is consistent with imaginary dielectric and electronic bandgap energy values. With the increase of photon energy, the absorption profile for all the considered perovskites is shown to be gradually increased in **Fig. 5(c)**. There are three major peaks observed from the photon energy-based absorption spectrum for the perovskites which are well-tuned with the complex dielectric functions' behavior. In the visible energy region (inset of **Fig. 5(c)**), doping by Ge in the B site with Sn leads to an increase in the magnitude of absorption coefficients than the pristine $CsSnI_3$ and $CsGeI_3$ compounds. Increasing the photon energy through the whole spectrum magnitude of the absorption profile also increases for all phases where the highest magnitude of $1.76e^6$ is found for $CsGeI_3$ perovskite. Ge-Sn mixed perovskites while leading in the visible spectrum, in the higher energy region they appeared to be dropping gradually with another peak showing at 17 eV. **Fig. 5(d)** shows that doping of Ge in B site with Sn leads to an increase in absorption in the whole visible spectrum wavelengths. Remarkably, the varied concentrations of Ge doping exhibit enhanced light absorption capabilities, causing specific perovskite formulations to outperform others. This opens intriguing possibilities for tailoring these perovskites for targeted applications in light detection and related life science applications. Notably, the 25% Ge doping concentration ($CsSn_{0.75}Ge_{0.25}I_3$) proves dominant across all compounds in the visible electromagnetic spectrum, particularly in absorbing yellow, orange, and blue light. Meanwhile, the 50% and 75% Ge-doped perovskites namely $CsSn_{0.50}Ge_{0.50}I_3$ and $CsSn_{0.50}Ge_{0.75}I_3$ demonstrate superiority in absorbing red, green, and violet light within the 380-400 nm region.

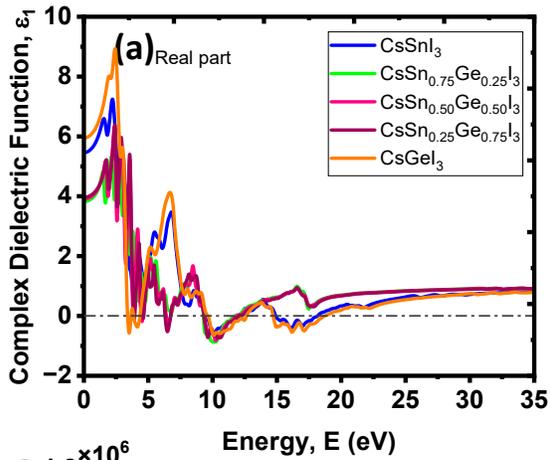
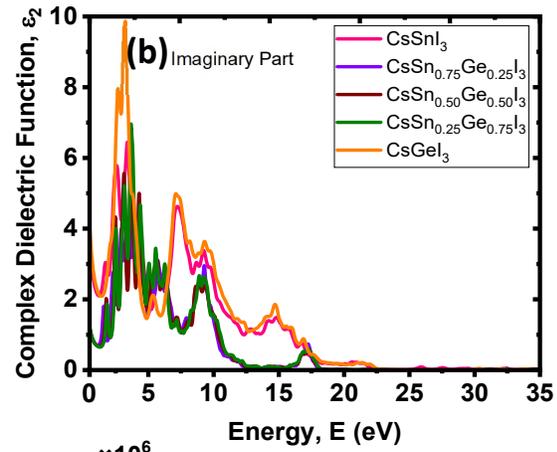
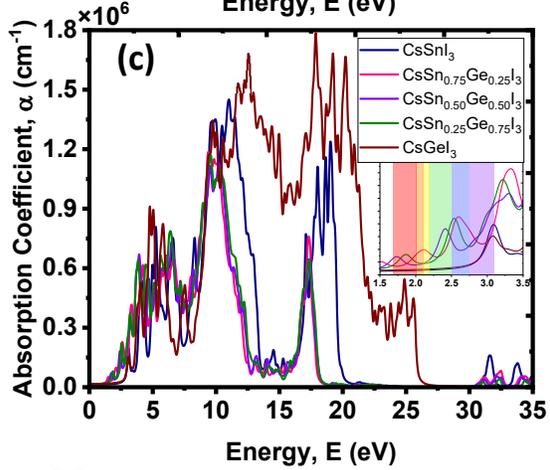
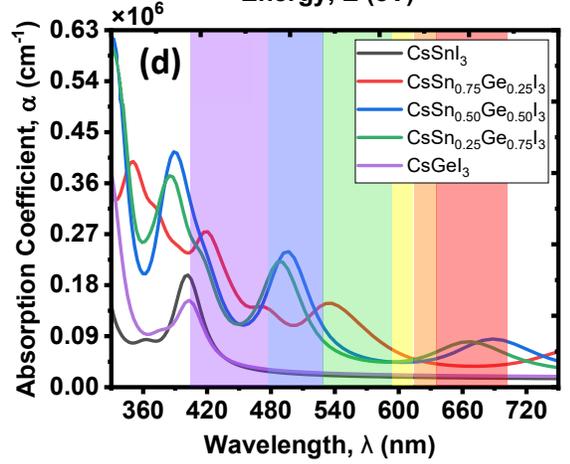
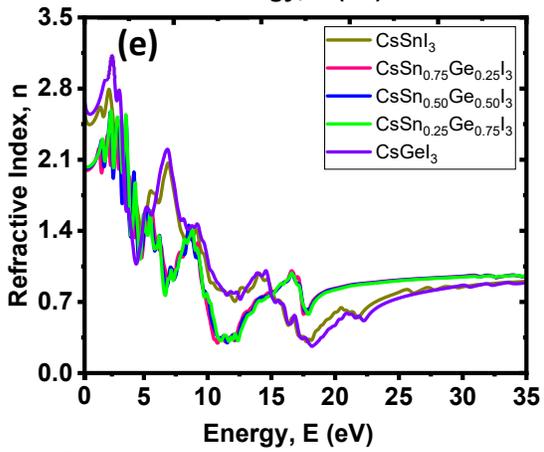
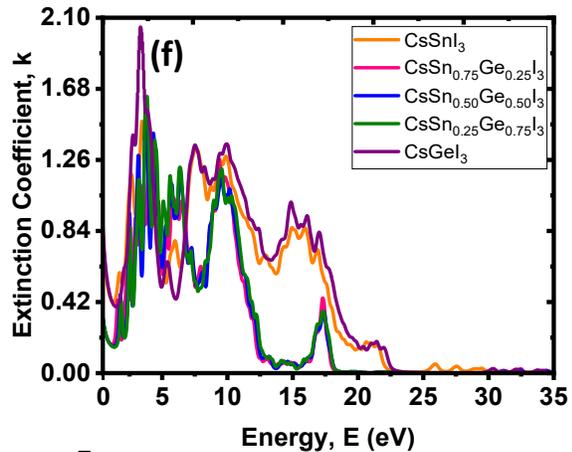
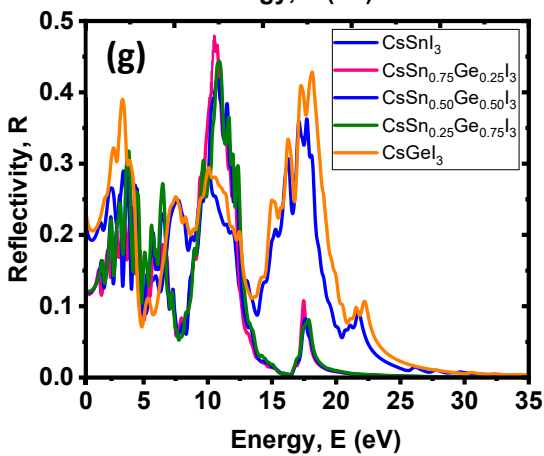
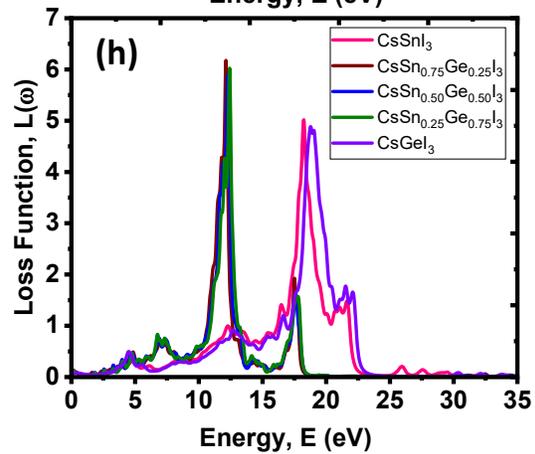

**Fig. 5.** HSE06 functional-based optical properties for: (a, b) Real and Imaginary Parts of Complex Dielectric Function (c, d) Photon Energy and Wavelength dependent Absorption Spectra, (e) Refractive Index n, (f) Extinction Coefficient k, (g) Reflectivity, and (h) Optical Loss Function for the considered perovskites varying with Ge concentration.

Complex refractive index describes the way of propagation of light through a medium. Though the values for the real part of the complex refractive index vary for the studied compound, the value of the imaginary part remains small for semiconducting materials. **Fig. 5(e, f)** represents the behavior of the complex refractive index values. In the real part of the refractive index from **Fig. 5(e)**, an almost qualitatively similar trend follows for the pristine perovskites and doped phases showing a decrement in magnitude values with increasing the photon energy. The static refractive index values are found in the range of 1.99 to 2.5 and upsurge to the highest value of 2.63 for Sn-free $CsGeI_3$ perovskite. In the visible energy region of photon energy, increment in x% concentration of Ge in $CsSnI_3$ demonstrates an increasing trend in both refractive index and extinction values as shown in **Fig. 5(e, f)**. Such higher values (>1) of the refractive index indicate their suitability as ambient optical materials for optoelectronic applications which are sufficiently greater than $SiO_2$ (1.45) to be potential for efficient energy harvesters from light in materials. In **Fig. 5(f)**, extinction coefficient profiles depict that a maximum value of 2.05 is observed for the $CsGeI_3$ perovskite which indicates a strong tendency to absorb incoming solar radiation. In the visible energy range, 25% doped compound $CsSn_{0.75}Ge_{0.25}I_3$ shows its dominance to lead by a higher magnitude than other doped phases. Considering the reflectivity as demonstrated in **Fig. 5(g)**, it shows that doping by Ge results in a drop of reflectivity for the $CsSnI_3$ perovskites where it increases with %Ge concentration increment but lower than the pristine compound. Static reflectivity R(0) values are found to be at 20%, 11%, 11.9%, 12.3%, and 21.8% for $CsSnI_3$, $CsSn_{0.75}Ge_{0.25}I_3$, $CsSn_{0.50}Ge_{0.50}I_3$, $CsSn_{0.25}Ge_{0.75}I_3$ and $CsGeI_3$ respectively. To determine the optical energy loss behavior with device applications, the optical loss function $L(\omega)$ is calculated and represented in **Fig. 5(h)** which describes systems loss in terms of dispersion, scattering, and heating effect. Photon energy dependent $L(\omega)$ exhibits that, in the visible energy region of the electromagnetic spectrum loss behavior is at the least magnitude whereas with higher photon energy loss function is increased to the highest by 6.17 at 12.1 eV for 25% Ge doped $CsSnI_3$ perovskite. This behavior shows that with higher energy than visible range radiative heat dissipation or uncontrolled scattering effect may affect the device efficiency. Moreover, pristine phases show loss below 1 until they reach to highest about 5 above 15 eV.

## DEVICE EFFICIENCY OPTIMIZATIONS:

### Bandgap and I-V Characteristics

Our DFT calculation shows that the bandgap of the absorber layer which is $CsSn_{1-x}Ge_xI_3$ (x = 0.00, 0.25, 0.50, 0.75, and 1.00) changes for different Ge doping levels in the perovskite material. DFT hybrid band structure calculation reveals that the bandgap of $CsSnGeI_3$ is 1.33 eV, 1.48 eV, 1.69 eV, 1.83 eV, and 1.92 eV for various Ge doping levels of x = 0.00, 0.25, 0.50, 0.75, and 1.00, respectively. The behavior of electrical parameters such as photocurrent and PCE is highly dependent on the energy band alignment of the absorber layer with ETL and HTL. **Fig. 6(a)** depicts the alignment of energy levels for the studied $CsSnGeI_3$ absorber, FTO, ETL, and HTL where 0%, 50%, and 100% levels of doping of Ge are considered in the perovskite material. 50% Ge doping in absorber material results in a bandgap of approximately 1.69 eV, which is ideal for electron-hole pair generation from visible light spectra. Whereas the ionization energy of HTL must be lower than that of $CsSn_{1-x}Ge_xI_3$ to extract the holes at the $CsSn_{1-x}Ge_xI_3$/HTL interface, the electron affinity of ETL must be larger than that of $CsSn_{1-x}Ge_xI_3$ to extract the electron at the ETL/ $CsSn_{1-x}Ge_xI_3$ interface [55].

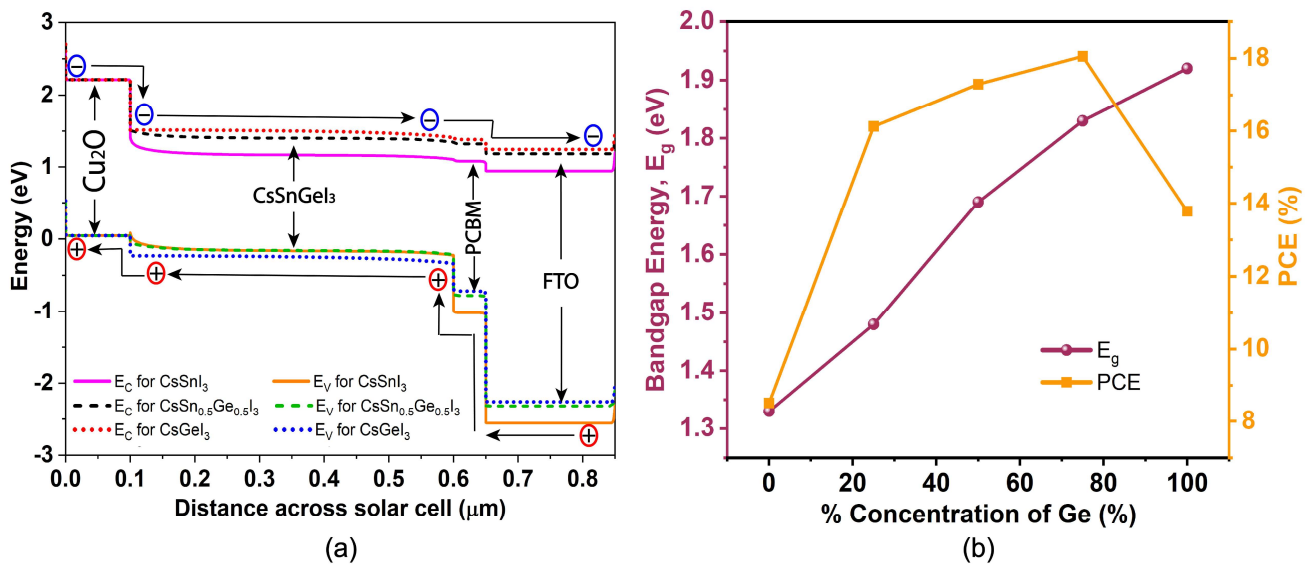

**Fig. 6.** (a) Perovskite Band Alignments with the doping concentration changes in B site of the $ABX_3$ phases. (b) Bandgap and PCE variation as a function of doping concentration of Ge.

The device's performance characteristics are also significantly impacted by the energy band imbalance at both interfaces. For all doping levels of Ge, **Fig. 6(a)** depicts an energy band escarpment at the conduction band (CB) of the HTL/absorber interface. The cliff is at its greatest for pure CsSnI$_3$ and is nearly identical for 50% and 100% Ge doping in CsSnI$_3$. While a lesser escarpment is advantageous, a larger deviation in band offset increases interfacial recombination, which degrades performance. The valence band (VB) cliffing is greatest also at the ETL/absorber interface for Ge without doping. So, the electrical efficacy of an absorber layer with no Ge doping will be poor. Alternatively, the cliffs at the ETL/perovskite interface in the VB and the HTL/perovskite interface in the CB for 50% (x = 0.50) and 100% (x = 1.0) doping are nearly identical, indicating that both doping levels will exhibit identical electrical performance. However, this is not the only factor that determines the electrical properties. As the absorber bandgap exceeds 1.80 eV for x% doping of Ge, it is unable to generate more electron-hole pairs from incident visible light. Our bandgap alignment diagram predicts that the 50% Ge doping level in the absorber material will result in better electrical properties than other doping levels of Ge. **Fig. 6(b)** depicts the changes in bandgap of CsSn$_{1-x}$Ge$_x$I$_3$ and PCE of the device as a function of doping levels of Ge. Bandgap of the absorber layer increases steadily as we increase the doping levels of Ge from 0.00% to 100% with a 25% step size. PCE of the device is comparatively low for 0% and 100% doping of Ge as these doping values withhold such bandgaps that are not suitable for absorbing lights in the visible range.

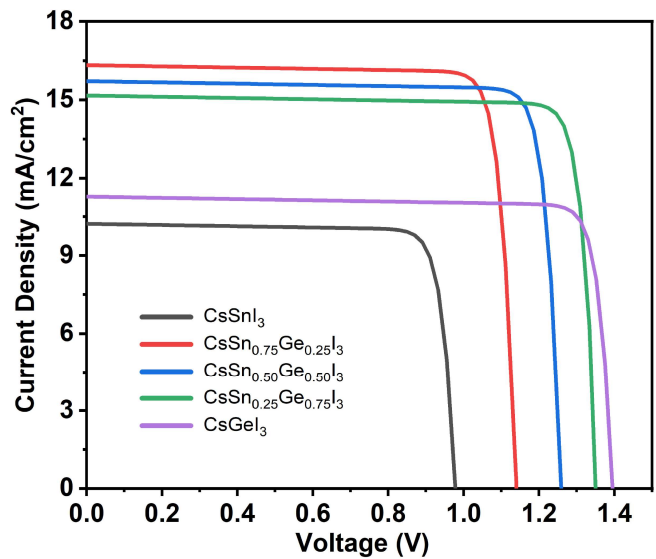

**Fig. 7.** I-V curves for different doping levels of Ge in CsSnI$_3$ perovskite.

One of the most crucial electrical properties of a solar cell is its current-voltage (I-V) curve, which is greatly influenced by the bandgap of the absorber material, the energy band alignment of the perovskite with the HTL and ETL, the absorption coefficient of the active layer, and the extinction coefficient of the perovskite material. **Fig. 7** illustrates the PSC's I-V curves for all Ge doping levels in CsSnGeI$_3$. Since the overall absorption coefficient for 25%, 50%, and 75% doping levels of Ge in the visible spectrum is the highest and nearly identical, we discovered the highest and identical J$_{SC}$ and PCE for these doping levels. **Fig. 5(d)** demonstrates that the absorption coefficient is low at Ge doping levels of 0% and 100%, which results in a decreased light-capturing capacity and low electron-hole pair creation. This occurrence is consistent with the lower photocurrent and PCE readings for these two doping levels as shown in **Fig. 7 & 8**. Understandably, the absorber layer with CsSnI$_3$ and CsGeI$_3$ should provide lower values of photocurrent and PCE as their absorption coefficients are lower in the visible range of light compared to other doping as evident in **Fig. 5 (d)** from HSE06 functional based optical properties.

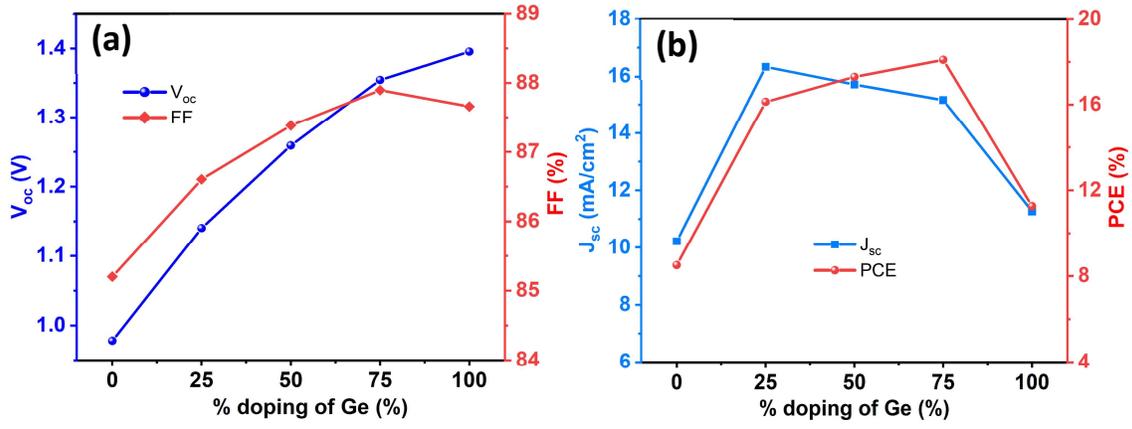

**Fig. 8.** (a) $V_{OC}$ and FF variation and (b) $J_{SC}$ and PCE curves due to the change in % doping concentration of Ge.

**Fig. 8(a)** shows the values of $V_{OC}$ and fill factor in relation to the amount of Ge doping in the PSC's active layer. As can be observed in **Fig. 8(a)**, the $V_{OC}$ rises as Ge doping levels in the absorber layer rise. If the bandgap is not too large, we typically see that the $V_{OC}$ increases as the active layer's bandgap advances. As the bandgaps are found to be larger for higher doping of Ge in $CsSn_{1-x}Ge_xI_3$, we found that the $V_{OC}$ also increases with the increment of doping levels of Ge. Understanding FF's increase with $V_{OC}$ is dependent on several factors, including Jsc and the maximum power point of the I-V curve. Because there is a reciprocal link between FF and $J_{SC}$, when the doping level of Ge is 100%, the $J_{SC}$ values fall significantly for that doping level, increasing FF. The I-V curve for 0% Ge doping level exhibits the lowest maximum power point of 140.5 W, which results in the lowest FF among all Ge doping levels since there is a proportional relationship between maximum power point and FF. **Fig. 8(b)** displays the values of $J_{SC}$ and PCE as a function of doping levels of Ge in the active layer. As previously observed from the hybrid HSE06 optical properties of DFT in **Fig. 5(d)**, the absorption coefficient is greater and nearly identical for 25%, 50%, and 75% doping levels of Ge, leading to higher and pretty similar Jsc values for these three doping levels. We have discovered that for active layers with poor absorption ability in pure $CsSnI_3$ and $CsGeI_3$ perovskites, the $J_{SC}$ values are likewise lower. This is because the $J_{SC}$ is significantly influenced by the amount of photogeneration occurring through absorption in the active layer. Following the same pattern as $J_{SC}$, we have noticed nearly the same trend in PCE for all the various doping levels of Ge.

*Effect of Bulk Defect Density*

The performance of the gadget is significantly influenced by the absorber layer's quality. A high recombination rate and low film quality can be caused by an absorber layer with a high defect density. Shockley–Read–Hall (SRH) recombination is the mechanism that can be used to describe the possible recombination techniques that have a direct effect on the performance of PSCs [56].

$$\mathcal{R}^{SRH} = \frac{V_{th}\sigma_n\sigma_p N_t[np-n_i^2]}{\sigma_n[n+n_1]+\sigma_p[p+p_1]} \quad (4)$$

$\sigma_n$ and $\sigma_p$ represent the capture cross sections of electrons and holes accordingly, $n_1$ and $p_1$ denote the trap defect density of electrons and holes respectively. $N_t$ represents the number of defects per volume. SRH recombination is proportional to the number of defects per volume. Increment in SRH recombination can degrade the performance of PSC in terms of efficiency. Defect density ($N_t$) has a direct influence on minority carrier lifetime ($\tau$). This influence can be understood by the following equation [56]:

$$\tau = \frac{1}{\sigma V_{th} N_t} \quad (5)$$

When defect density decreases, the carrier lifespan increases ensuring better solar cell performance.

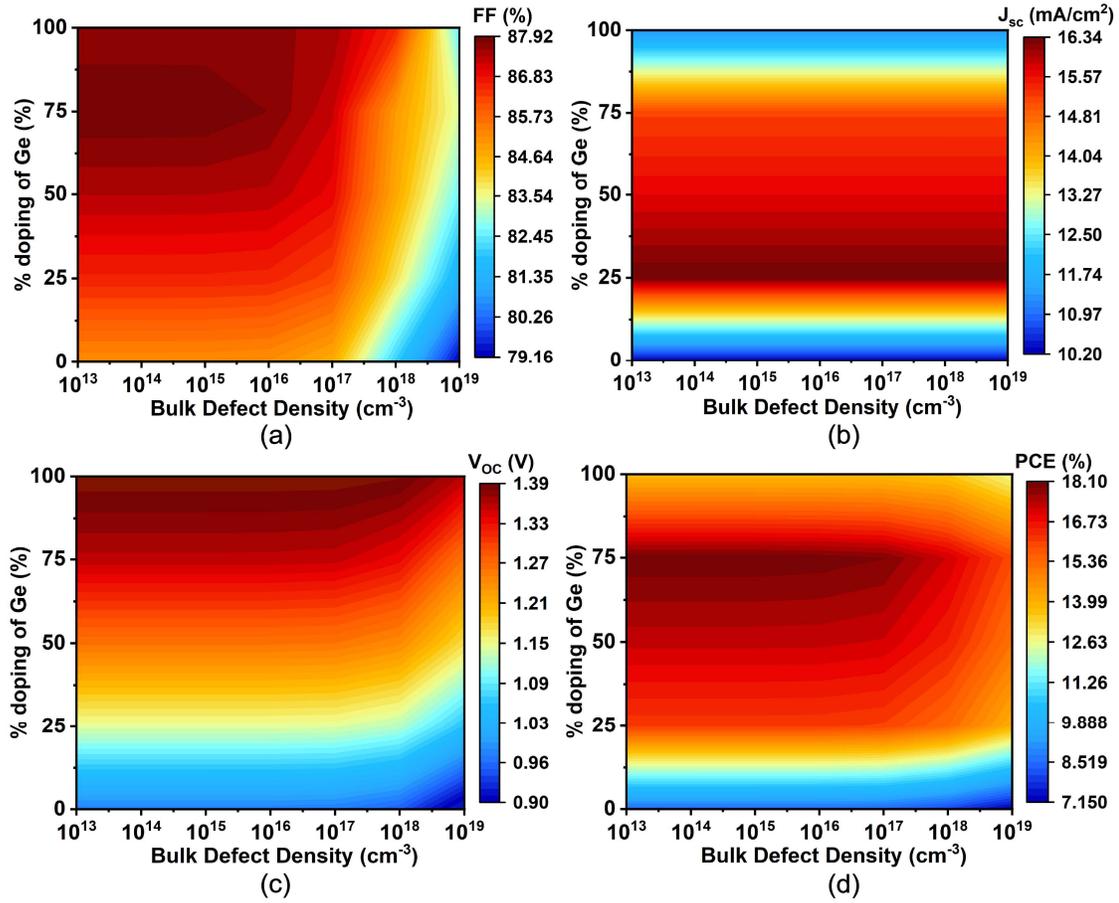

**Fig. 9.** Effect of bulk defect for $CsSn_{1-x}Ge_xI_3$ (where x = 0.00:0.25:1.00) on (a) FF, (b) $J_{SC}$, (c) $V_{OC}$ and (d) PCE.

**Fig. 9(a)** displays the color mapping of FF for various Ge doping concentrations and absorber layer bulk defect densities. Compared to other electrical characteristics, FF is more sensitive to defect density. FF diminishes when the defect density increases significantly. This is because when defect density rises significantly, the carrier lifetime in the absorber layer falls, resulting in a decrease in the photocurrent generated inside the solar cell [57]. **Fig. 9(c)** shows the color mapping of Voc for different levels of doping in Ge with different densities of bulk defects in the absorber layer. We have seen a similar trend in the contour plot when the Voc increases for large Ge doping levels in the absorber, as was stated before. Even if there is 100% doping of Ge in the absorber layer, $V_{oc}$ drops as the defect increases excessively. As shown in **Fig. 9(b)**, the defect density has no significant effect on $J_{sc}$ for any doping level of Ge in the absorber layer. Maximum photocurrent production of a PSC depends more on the active layer's capacity for absorption than it does on the density of nearby defects. The maximum PCE for Ge doping levels of 50% and 75% have been observed in the early analysis, and **Fig. 9 (d)** confirms this.

Efficiency decreases when defect density rises as a result of shorter carrier lifetimes and higher SRH recombination rates in the absorber layer.

*Impact of Absorber Thickness*

The performance of a solar cell is largely dependent on the thickness of the perovskite layer ($t_{perov}$). While keeping the thickness of the other layers the same, the thickness of the absorber layer is changed from 200 nm to 800 nm to study this impact. The rate of recombination and extraction of charge carriers is significantly impacted when $t_{perov}$ is changed because of the relationship between $t_{perov}$ and the diffusion length of the light carriers. The performance of the device will suffer if the absorber layer is made extremely thick since this would cause the charge carriers to recombine before they reach the electrodes, which will be counterproductive. Nevertheless, the photovoltaic performance of the device would suffer if the absorber layer is too thin, as a result of a decrease in the quantity of photogenerated charge carriers [58], [59]. **Fig. 10 (c)** shows the Voc values as a function of absorber layer thickness from 200 nm to 800 nm for a variety of Ge doping levels. Larger absorber thicknesses simply boost photocurrent production; therefore, they have no significant effect on open circuit voltages. FF values almost follow the same pattern as Voc values as illustrated in **Fig. 10 (a)**. However, FF values are found to be a little bit more sensitive for CsGeI$_3$. The probable reason is that when the absorber layer thickness gets too thin, it generates a very small amount of photocurrent in addition to having much less absorption ability for 100% doping of Ge. According to **Fig. 10 (b)**, the Jsc steadily increases for all cases as the thickness grows from 200 nm to 800 nm. The concentration of excess carriers grows as the layer's thickness does, leading to an increase in Jsc. Due to the proportional relationship between Jsc and PCE, PCE exhibits the same pattern as Jsc values as absorber thickness is raised as depicted in **Fig. 10 (d)**.

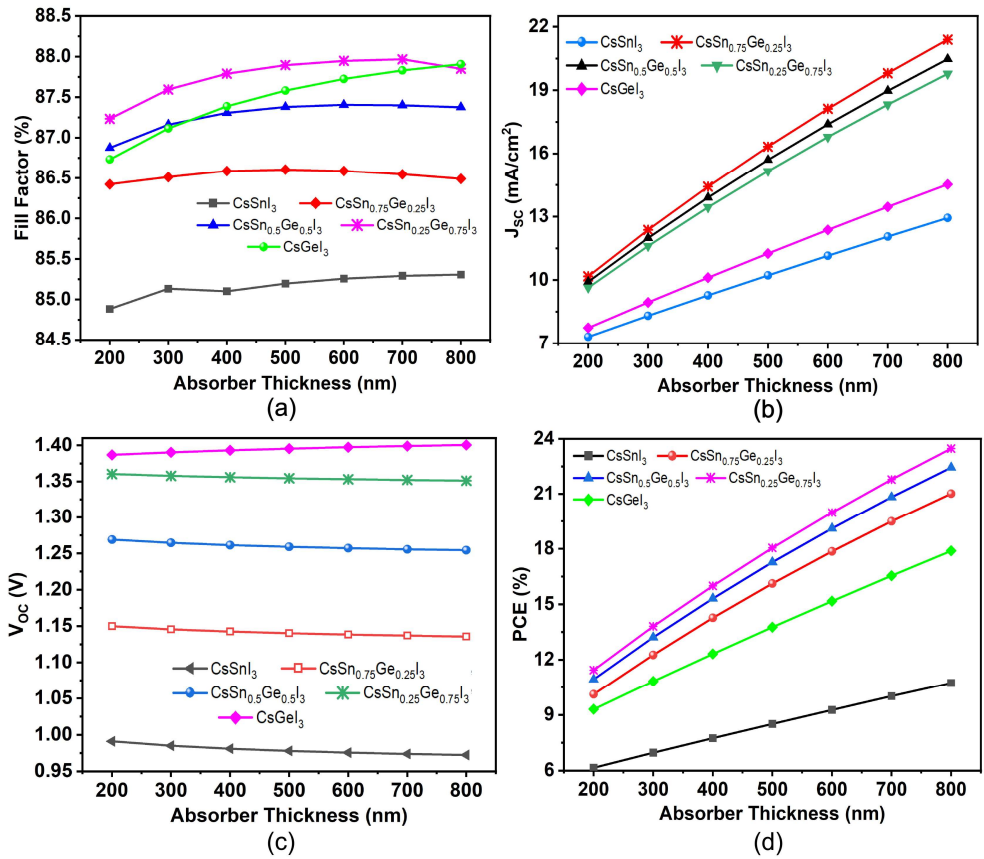

**Fig. 10.** Effect of absorber thickness for CsSn$_{1-x}$Ge$_x$I$_3$ (where x = 0, 0.25, 0.50, 0.75, and 1.00) on (a) FF, (b) J$_{SC}$, (c) V$_{OC}$ and (d) PCE.

*Analysis of Electric Field Distribution*

When it comes to the performance analysis and justification of the generating spectra, the distribution of the electric field throughout the solar cell is an essential component. The electric field is computed at five distinct incoming light wavelengths: 300 nm, 550 nm, 705 nm, 1320 nm, and 2000 nm.

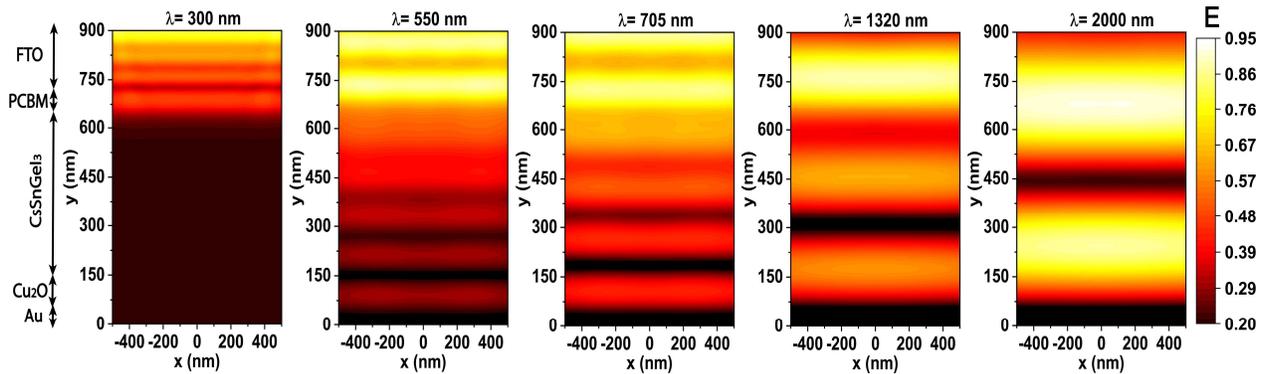

**Fig. 11.** Electric field distributions for five different wavelengths of incident light for the solar cell with 50% ($CsSn_{0.75}Ge_{0.25}I_3$) doping of Ge

As illustrated in **Fig. 11**, the maximum E field intensity is observed towards the top layer of PSC when the wavelength of the light is low. This is because short-wavelength light can only penetrate to a shallow depth, and the ETL layer possesses a high absorption coefficient [60]. The maximal electric field intensity broadens sufficiently to approach the ETL/perovskite interface when the wavelength is extended to 550 nm as depicted in **Fig. 11**. The solar cell layers are easily penetrated from top to bottom by the larger wavelengths. When the wavelength of the light extends above 700 nm, the Fabry-Perrot resonance induced by the gold back reflector causes the electric field to produce a standing wave within the perovskite layer. As shown in **Fig. 11**, the rear metal contact of the solar cell is sufficiently thick to reflect the majority of the light that passes through it, and the high absorptivity of the perovskite allows it to absorb the vast majority of the light. As 50-75% Ge doping in the absorber layer are the best doping levels for absorption among others, high photocurrent generation, and high PCE, the whole electric field calculation is performed for this range of doping levels.

### *Effect of Doping Levels of Ge on Generation Rate*

The generation rate is found to be greatest at the interface of $CsSn_{1-x}Ge_xI_3$ and PCBM due to the potential for recurrent reflection of light and electron-hole pair production within the interface. The majority of the incident light is absorbed in the ETL/absorber contact, which causes the generation rate to decrease as we approach further into the active region of the cell. A solar cell's generation rate may be expressed as the volume of the device divided by the total number of electrons that are created by the cell. We ran simulations with 0%, 50%, and 100% doping of Ge to examine the impact of Ge doping levels in $CsSn_{1-x}Ge_xI_3$ on the generation rate as illustrated in **Fig. 12 (b)** Due to the high absorption coefficient and low reflectivity of $CsSn_{0.50}Ge_{0.50}I_3$, which produces a greater number of electron-hole pairs, the generation rate at the interface is greatest for a Ge doping level of 50% near 630 nm along the Y axis of the cell.

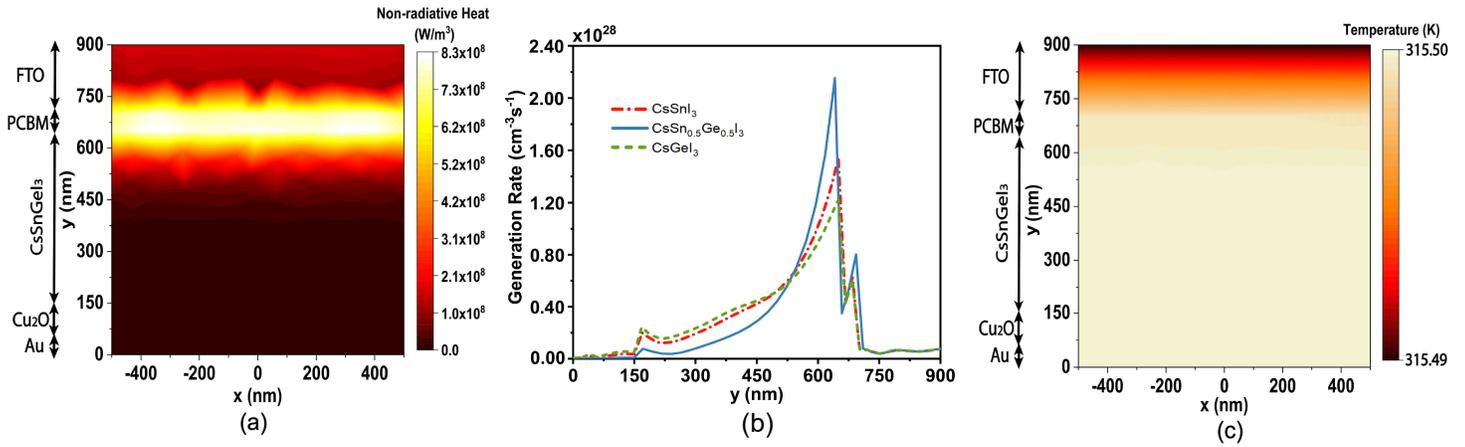

**Fig. 12.** (a) Heat Profile, (b) Generation Rate, and (c) Temperature distribution across the perovskite device.

*Heat and Thermal Analysis*

The thermal simulation is run to understand the distribution of produced non-radiative heat in the solar cell structures and the temperature gradient in each layer by connecting the thermal module to the optic module of the simulator. **Fig. 12 (a)** exhibits the spatial pattern of the non-radiative heat produced by the PSC with 50% Ge doping levels in the absorber layer. The SRH recombination rate is typically at its highest point near the ETL/perovskite interface, which results in a significant increase in heat between 600 nm and 750 nm along the Y axis of the PSC. The thermalization process, which is spatially dependent on the generation rate profile, also uses the surplus energy over the bandgap as a source of heat creation [61]. As we have observed that the generation rate is greatest at the ETL/perovskite interface, we have discovered that this interface also produces the most non-radiative heat as seen in **Fig. 12 (a)**.

The solar cell's ambient temperature was recorded at 300K before solar generation while the simulation was running. As seen in **Fig. 12 (c)**, the PSC temperature rises by roughly 15.52K following solar generation. Because the solar cell is so thin, there is hardly any temperature variation from top to bottom. Because transparent glass has a low heat absorptivity, the top layer of the PSC is somewhat cooler than the other layers, according to the temperature distribution. It is important to note that this temperature map is intended for a steady-state calculation and that a time-dependent computation will still need to take into account the dynamic change of the temperature during iteration.

## Summary and Perspective

Our multimodal numerical investigation reveals that by selectively varying the absorber layers in perovskite solar cells while maintaining standard electron and hole transport layers, the device efficiency can be significantly improved. Our findings emphasize the importance of doping in the perovskite layer to achieve optimal power conversion efficiency (PCE) in solar cell devices. We demonstrate the use of density functional theory (DFT) to tune the physical light-matter interaction in $CsSnI_3$ perovskite materials through atomic doping of Ge. The DFT calculations indicate that increasing the concentration of Ge leads to an increased bandgap and enhanced absorption profile, resulting in improved solar energy conversion efficiency. Through our study, we illustrate how perovskites can dynamically tune device parameters and interact with light based on their structural dynamics and electronic properties. Furthermore, we demonstrate that optimizing the absorber thickness in conjunction with doping concentration variations can further enhance device efficiency. Our optimized $CsSnI_3$ PSC with a 75% Ge concentration ($CsSn_{0.25}Ge_{0.75}I_3$) achieves a remarkable power conversion efficiency of 23.8%, accompanied by a Voc of 1.40 V, Jsc of 21.56 mA/cm$^2$, and FF of 87.80%. By carefully tuning the perovskite absorber layer composition, we successfully elevate the PCE of standalone Sn-based cubic iodide perovskite from 10.5% to 23.8%. Additionally, thermal and field distribution studies for the considered perovskite-based devices in this study highlight the significant impact of doping in the absorber layer on exciton dynamics in practical device applications. This comprehensive investigation, focusing on tuning the photo-physics of the perovskite absorber layer through atomic doping, instead of other mechanisms, opens up new possibilities for the development of low-cost, high-efficiency solar cells with improved stability.

**Declaration of competing interest**

The authors declare that they have no known competing financial interests or personal relationships that could have appeared to influence the work reported in this paper.

**Data availability**

Data will be made available on request.